# Optimizing the Ptolemaic Model of Planetary and Solar Motion

Ilia Rushkin

## Abstract

Ptolemy's planetary model is an ancient geocentric astronomical model, describing the observed motion of the Sun and the planets. Ptolemy accounted for the deviations of planetary orbits from perfect circles by introducing two small and equal shifts into his model. We show that Ptolemy's choice of shifts allowed him to approximate the true eccentricity of planetary orbits only in the first order, linear in eccentricity. We show that, if the shifts were tuned in the ratio 5/3, the model's precision could be improved substantially, perhaps delaying its rejection. The best achievable precision is quadratic in eccentricities. Although, to achieve it fully, one would have to introduce Ptolemaic shifts in the epicycles as well as in deferents the planets.

## Introduction

It is common knowledge that the ancients assumed that the Sun and other planets rotate around Earth, and that only relatively recently Copernicus and others threw the geocentric model out and replaced it with the correct heliocentric one. Actually, this is one of those stories whose short version is much more straightforward than the full version. Historical reality was somewhat more complicated. It is true, the astronomers of the ages before Copernicus or Galileo did not know the words "changing the reference frame", but they understood the idea sufficiently well. Mathematically, they were much more comfortable with it than most people think. Nor was there a consensus among the ancients about a single geocentric model. Our knowledge of the variety of ancient celestial models in very sketchy, but we know, for example, that the Greeks also considered combined models, in which at least some of the planets rotate around the Sun, while the Sun rotates around Earth.

In the second century A.D., Alexandrian polymath Claudius Ptolemy produced a large astronomical treatise, known to us under the Arabized name of the Almagest. This book was a sum total of the Greek knowledge (which incorporated the knowledge of other Near Eastern civilizations) about planetary motion, and its impact on the posterity proved to be enormous. The specific version of the geocentric model, posited in this book, was widely accepted during the remainder of antiquity and all of the Middle Ages, until seventeenth century. It was this book that Copernicus and Kepler were disproving. Below, we will always have this book in mind, subsuming the variety of ancient astronomers under a single name "Ptolemy". It may seem like crediting him with the achievements of others, but otherwise this paper would quickly turn into a historical thesis.

Most modern people are used to disparaging the geocentric astronomy. Similar to the assumption of flat Earth, it is regarded as a symbol of ignorance. True, Ptolemy's model of the Solar system turned out to be wrong, just as most other ancient theories. But this should not prevent us from appreciating what a huge intellectual achievement this model was. Besides, the ancient observational data were limited and low-precision, which meant that for a very long time Ptolemy's model adequately described and allowed predicting most (though not all) of the available observations. The general methodology of science forces us, therefore, to admit that there was nothing ridiculous in this model. For its time it was rather good.

We will briefly describe Ptolemy's geocentric model in modern terms and show that there was room in it for improvement. If all the planetary orbits (here we mean the "true" Newtonian ones) were perfect circles, Ptolemy's system could describe the observations perfectly. The discrepancies arise because of the orbit eccentricities, which are all small, and so one can include them into the model via a perturbation series. We will show that Ptolemy tuned the parameters of his model so as to account for the first term of the series (linear in eccentricities), but that it was possible, by tuning them differently, to account also for the second term (quadratic in eccentricities). Had Ptolemy done it, the model would be capable of describing the observations much better, perhaps postponing the invention of Kepler's elliptical orbits by some decades.

## Ptolemy's model of planets and the Sun

We will compare Ptolemy's geocentric model with the classical modern model of the Solar system, in which the Sun is at rest and the planets move around it in coplanar elliptical orbits according to Newton's law of gravity, neglecting all interactions except the attraction of each planet to the Sun. Such motion can also be called "Newtonian". Even though this model has obvious limitations, for the sake of brevity we will often refer to this model simply as "reality" or "true".

## Zeroth-order approximation: round orbits

Let us first consider the Solar system in the zeroth-order approximation, neglecting all the orbital eccentricities entirely. Then, in the true model, all the planets move around the Sun in perfect circles with constant angular velocities. Switch to the reference frame in which Earth is at rest. This means simply subtracting Earth's orbital motion from the motion of each other planet (or the Sun). Hence, as observed from Earth, the motion of a planet will be a sum of two circular motions. It moves around a circle, whose center itself moves around another circle, in whose center Earth resides.

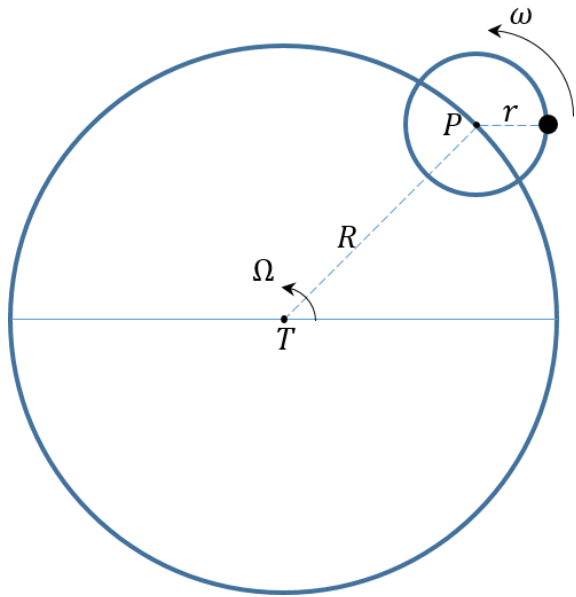

**Figure 1.** The basic Ptolemy's model for the motion of a single planet. Eccentricity is ignored. The position of the observer (Earth) is marked $T$. If this planet is Mercury or Venus, then from the modern point of view the Sun should be placed in the point $P$. Ptolemy, who was unable to measure distances, knew that the Sun lies somewhere along the same radius as $P$.

This is Ptolemy's basic idea of planetary motion. The moving circle is called the *epicycle* ("the add-on circle"), and the stationary circle is called the *deferent*[1] ("the carrying circle"). The angular velocities of both circular motions are constants. For the Sun, the radius of the epicycle is zero. In other words, it has no epicycle and moves along its deferent. Note that Ptolemy measures all motions in the reference frame of fixed stars, therefore the diurnal spinning of Earth is already taken into account and we do not have to worry about its existence. We should also specify that Ptolemy had no way of measuring distances to the celestial bodies: precise measurement of the angular sizes of the planets and the Sun was out of his reach. It is important to appreciate that he had access to *angular* data only. Thus, for each body, he knew the ratio of the epicycle radius to the deferent radius, but he had no way of finding them out separately. As a result, Ptolemy strove to create a mathematical model that correctly described the observed angles and angular velocities, but he had no idea whether it described distances well.[2]

---

[1] Of course, Ptolemy wrote in Greek. But as in other European sciences, Greek was supplanted by Latin, and so we use the conventional Anglicized version of Latin terminology.

Of course, decomposing the motion into a sum of two circular motions is entirely commutative. We could switch the deferent and the epicycle, and the net motion of the planet would be the same. The Greeks followed a simple rule: the larger of the two circles is the deferent, and the smaller is the epicycle. This convention, though natural, causes unfortunate complications from the modern standpoint of comparing it to the real Solar system. For Mercury and Venus – the *inferior* planets, i.e. those with true orbits within Earth's true orbit, – the epicycle is the planet's true motion, and the deferent is the Earth's true motion with a minus sign. But for the other, *superior* planets (of which Ptolemy knew three: Mars, Jupiter and Saturn) the roles are reversed: the epicycle is Earth's true motion, and the deferent is planet's true motion. Therefore, for Mercury and Venus $\Omega = \frac{2\pi}{1 \text{ year}}$, and the angular phases of their deferent motions are both the same as that of the Sun. But their $\omega$'s are planet-specific. For Mars, Jupiter and Saturn the situation is reversed: $\Omega$'s are planet-specific, but $\omega = \frac{2\pi}{1 \text{ year}}$ for all three, and the angular phases of their epicycle motions are all the same as that of the Sun[3].

## Accounting for orbital eccentricity in Ptolemy's model

If all the true orbits of the planets had zero eccentricity, Ptolemy's model would work perfectly. But the precision of ancient observations was sufficient to see beyond this approximation, and therefore Ptolemy had to modify the model to account for deviations from the simple circular motion. In retrospect, we know that the "correct" modification would have been to make both the epicycle and the deferent ellipses rather than circles, and implement Kepler's second law instead of assuming constant $\Omega$ and $\omega$. Then the model would be correct: it would differ from the Newtonian model only in that it uses a different reference frame.

---

[2] The only sad exception was the Moon. To fit the data, Ptolemy had to introduce for it an epicycle mechanism, similar and even more complex than for a planet. But this model predicted that the distance to the Moon varies almost by a factor of 2, and so should its visible size, which is clearly at odds with the reality. This difficulty of Ptolemy's theory was recognized and never resolved.

[3] By the way, it is precisely this observation that led Copernicus to his heliocentric model. He decided that there are too many coincidences in Ptolemy's model and recognized that all of them can be explained away by switching to the reference frame of the Sun. Copernicus was driven not by new observations or mathematical techniques (in his model the orbits were circular, and epicycles were still present), but simply by the desire for elegance.

But Ptolemy chose a different path. He made no changes at all in the motion of a planet in the epicycle, but introduced changes in the motion of the epicycle's center around the deferent. Firstly, he kept the deferent's shape a circle, but shifted its center $C$ away from Earth. The ratio $\frac{TC}{R}$ is traditionally called the eccentricity, but we will avoid this so as not to confuse it with the true eccentricity of elliptical planetary orbits (if one compares the models, Ptolemy's eccentricity coincides with the real one in the leading order, when both are small). Secondly, he described the deviation from constant rate of rotation by introducing a third point $E$, called the *equant*. Ptolemy postulated that the speed of motion of $P$ is such, that the rotation of the radius $EP$ has constant angular velocity $\Omega$.

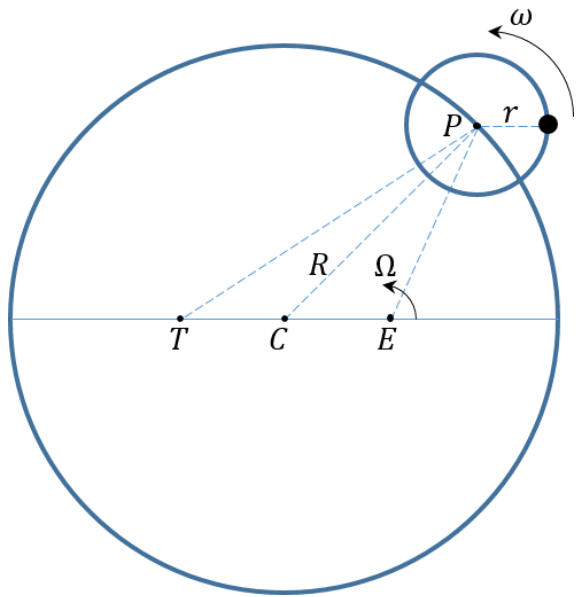

**Figure 2.** Ptolemy's model of motion for a single planet. The sizes of the shifts $TC$ (from Earth to the deferent center) and $CE$ (from the deferent center to the equant) are greatly exaggerated.

Probably, this was not an easy compromise for a Greek, who saw a special philosophical significance in perfect circles and uniform motion. The center of the epicycle still moves in a circle, although a shifted one; and it still moves with a constant angular speed, although only if viewed from a point shifted from the center.

Now Ptolemy had some free parameters to tune to approximate the observational data. For each planet, using its deferent radius as the length unit, he found the epicycle radius (0 for the Sun), and also the shifts. And here he made some assumptions. For the sake of simplicity and not having sufficiently clear data to decide against it, Ptolemy assumed that $TC = CE$ for any planet, whereas for the Sun $CE = 0$.

These assumptions are suboptimal. We will show that, using them, it is theoretically possible to account for non-circularity of true orbital motions to the first order in eccentricities[4]. But this is not the best that this model is capable of. Relaxing Ptolemy's assumptions, it is possible to get to the second order. This is the "true ceiling" of the model: to account for ellipticity in the third order and higher one would have to introduce new parameters.

To be exact, in order to achieve the precision quadratic both in the eccentricity of the true orbit of the planet and the eccentricity of the true orbit of Earth, we would need to introduce shifts in the epicycle in the same way as Ptolemy did in the deferent. But this extra modification would not be too big a crime in Ptolemy's eyes. He did it once, so why not do it again? Secondly, even if we don't introduce shifts in the epicycle and only tune Ptolemy's shifts differently, we could improve the model's precision significantly. The fact is, the true eccentricity of Venus is just a fraction of the eccentricity of Earth's orbit, which is in turn just a fraction of the eccentricities of Mars, Saturn and Jupiter. The planets conspire in such a way that the true motion, which is to be approximated by the motion in Ptolemy's epicycle, is not only smaller, but also significantly more circular than the motion in the deferent. This is precisely why Ptolemy could "get away" with perturbing the deferent but leaving the epicycle as is. Mercury is the sole and notable exception to this happy rule. Ptolemy's theory for it is much worse than for other planets, and it is easy for us to understand why. Mercury's true orbit has an abnormally high eccentricity 0.206, well outside the comfort zone for the small-eccentricity description as such, let alone the intricacies of motion decomposition within that approximation. In addition, being close to the Sun, Mercury is difficult to observe, so Ptolemy had especially poor data for it.

---

[4] How well Ptolemy managed to achieve this precision is another matter. He had to work with rather imperfect observational data.

Given how Ptolemy's deferents and epicycles correspond with the true planetary motions, it is possible to re-calculate the parameters of true orbital motion from Ptolemy's parameters in the Almagest (see [1] for a comprehensive list of Ptolemy's values). The table below shows the results of such calculations and compares them with the modern values to give the reader an idea of Ptolemy's overall level of precision.

| Object | Eccentricity (planet's orbit for superior planets, Earth's orbit otherwise) | | | Planetary orbit's major semi-axis in astronomical units | | | Planetary orbital period in years | | |
|---|---|---|---|---|---|---|---|---|---|
| | Ptolemy | True value | Match | Ptolemy | True value | Match | Ptolemy | True value | Match |
| The Sun | 0.0208 | 0.0167 | 125% | | | | | | |
| Mercury | 0.0500 | 0.0167 | 299% | 0.375 | 0.387 | 97% | 0.241 | 0.241 | 100% |
| Venus | 0.0208 | 0.0167 | 125% | 0.719 | 0.723 | 99% | 0.615 | 0.615 | 100% |
| Mars | 0.1000 | 0.0943 | 106% | 1.519 | 1.524 | 100% | 1.881 | 1.881 | 100% |
| Jupiter | 0.0458 | 0.0488 | 94% | 5.217 | 5.204 | 100% | 11.862 | 11.862 | 100% |
| Saturn | 0.0569 | 0.0557 | 102% | 9.231 | 9.582 | 96% | 29.487 | 29.457 | 100% |

The estimates for the planetary periods are unsurprisingly excellent. The estimates for the orbit sizes are good. The estimates for eccentricity are good for the superior planets (Mars, Jupiter, Saturn), and bad for Earth. Especially drastic is the overestimate of Earth's eccentricity in case of Mercury. As we already mentioned, Mercury is particularly "difficult" planet for Ptolemy. It may seem surprising that the eccentricity of Earth's orbit is so crude in Ptolemy's theory of the Sun, for it may seem that this theory is the simplest one: there is no epicycle, and observations of the Sun should have been abundant. However, let us not forget that the measurements of the Sun's position were different from other planets (those could be measured *relative* to the Sun), and that tuning the parameters for the Sun Ptolemy had other overriding concerns, such as getting the equinox and solstice dates right. The overestimate of Earth's orbital eccentricity in case of Venus is in excellent agreement with the case of the Sun: one is a consequence of the other.

## Solution of the general Ptolemy's model

Let us generalize Ptolemy's model a little: the lengths of $TC$ and $CE$ do not have to be the same. In principle, we could also make their directions independent. Symmetry considerations suggest that this is pointless for approximating the motion of planets, so we will not do it. But if we wanted to introduce a non-zero angle $\phi$ between the two shifts, the modification would be simply changing $\alpha \to \alpha + \phi$ in the equations.

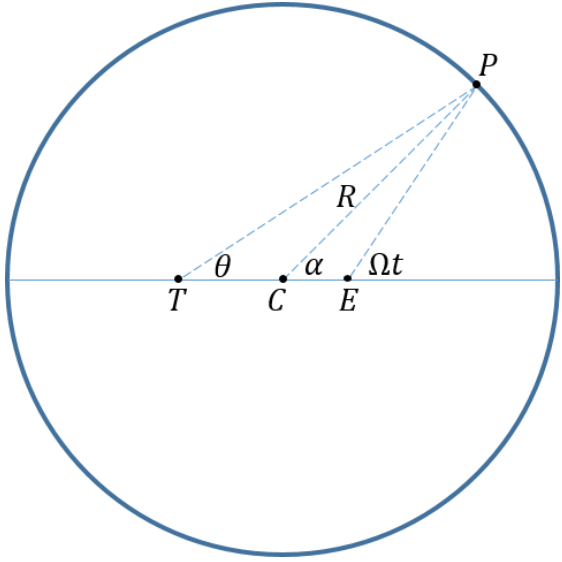

**Figure 3.** Generalized Ptolemy's model, where $TC$ and $CE$ are not necessarily equal. Only the motion of the center of the epicycle is shown.

The main problem is to find the time-dependence of the azimuthal angle $\alpha$, as seen from the circle center $C$. A simple way to do it is to recall the law of sines from geometry: in a triangle, the ratio of the sine of any angle to the length of the opposite side is constant. Apply this theorem to the triangle $CPE$, while noting that angle $\angle CPE = \Omega t - \alpha$, and that $\sin \angle CEP = \sin(\Omega t)$. We get $\sin(\Omega t - \alpha) = \frac{CE}{R}\sin(\Omega t)$. From here

$$\alpha(t) = \Omega t - \sin^{-1}\left(\frac{CE}{R}\sin(\Omega t)\right). \tag{1}$$

Now let's find the time-dependence of the azimuthal angle $\theta$, as seen from Earth. Apply the law of sines to the triangle $TPC$: $\sin(\alpha - \theta) = \sin\alpha \cos\theta - \cos\alpha \sin\theta = \frac{TC}{R}\sin\theta$,

$$\theta(t) = \tan^{-1}\frac{\sin\alpha(t)}{\cos\alpha(t) + \frac{TC}{R}} \tag{2}$$

If we wish to express the time-dependence of $\theta$ explicitly, we can substitute here the formula (1) for $\alpha$. Finally, the distance $TP$ from Earth to the planet is found from the law of sines again, applied to the triangle $TPC$:

$$TP = R\frac{\sin \alpha(t)}{\sin \theta(t)}. \qquad (3)$$

# Optimizing Ptolemy's model

Let us examine what shifts $TC$ and $CE$ produce the closest approximation to the Newtonian planetary motion (in case of Ptolemy's theories of superior planets – the motion of those planets; in case of his theories of inferior planets – the motion of Earth). Recall that it is an approximation in which distances are ignored, because they are not observable to Ptolemy. Only angles and angular speeds are observable, so we need to consider those. In Newtonian mechanics, for a planet in an elliptical orbital motion,

$$\frac{d\theta}{dt} = \frac{\Omega}{(1-e^2)^{\frac{3}{2}}}(1-e\cos\theta)^2 \qquad (4)$$

where $e$ is the eccentricity of the orbit. Let us look at the corresponding expression in Ptolemy's model and ask: which values of the shifts $CE$ and $TC$ will produce the closest match of the two models in the limit of small eccentricity? From general considerations, we expect that in this limit the ratios $\frac{CE}{R}$ and $\frac{TC}{R}$ are linear in eccentricity $e$, and that is indeed what we will find.

A simple way to do it is to go backwards: use the same equations as before, but express $\Omega t$ as a function of $\theta$ rather than vice versa. We get:

$$\tan(\Omega t) = \frac{\sin\left(\theta + \sin^{-1}\left(\frac{TC}{R}\sin\theta\right)\right)}{\cos\left(\theta + \sin^{-1}\left(\frac{TC}{R}\sin\theta\right)\right) - \frac{CE}{R}} \qquad (5)$$

Expand to the second order in the shift ratios:

$$\tan(\Omega t) \approx \frac{\sin\left(\theta + \frac{TC}{R}\sin\theta\right)}{\cos\left(\theta + \frac{TC}{R}\sin\theta\right) - \frac{CE}{R}} \qquad (6)$$

Differentiate both sides with respect to time using the identity $\tan' x = 1 + \tan^2 x$:

$$\frac{d\theta}{dt} \approx \Omega \left( 1 + \left(\frac{CE}{R}\right)^2 + \frac{TC}{R} \cdot \frac{CE}{R} - \left(\frac{CE}{R} + \frac{TC}{R}\right) \cos\theta + \left(\left(\frac{TC}{R}\right)^2 - \left(\frac{CE}{R}\right)^2\right) \cos^2\theta \right) \tag{7}$$

Now let us equate the zeroth and the first harmonics between the two models (Eq. 4 and Eq. 7), staying at the same precision:

$$2e^2 = \frac{1}{2}\left(\frac{CE}{R} + \frac{TC}{R}\right)^2$$

$$2e = \frac{CE}{R} + \frac{TC}{R}$$

We see that these equations are satisfied together and therefore give just one constraint on the choice of the shifts. This is how far Ptolemy effectively got: the sum of his two shifts equals twice the eccentricity (witness the reasonable matches in the table in the previous section), but he had no idea about the best way to divide this distance between the two shifts. He assumed that, in case of the Sun, $TC$ gets everything and $CE$ is zero, whereas in case of a planet the two shifts are equal. We, however, can proceed and demand the equality of the second harmonics in both models:

$$e^2 = \left(\frac{TC}{R}\right)^2 - \left(\frac{CE}{R}\right)^2$$

Immediately, we see that Ptolemy's choice of shifts does not work. The optimal parameters are:

$$\frac{TC}{R} = \frac{5}{4}e, \quad \frac{CE}{R} = \frac{3}{4}e, \quad \text{meaning that} \quad \frac{TC}{CE} = \frac{5}{3} \tag{8}$$

We obtained these values from two equations using two unknowns, so we have run out of variables. This means that to achieve third order and higher precision in eccentricity one would have to introduce more variable parameters into Ptolemy's model.

One might feel that, since harmonic analysis was unknown to Ptolemy, it is somehow unfair to use it as a criterion for optimizing his model. Therefore, it is worth noting a different and very simple method:

equating the minimal and maximal values of the angular velocity from both models, which naturally occur at $\theta = 0$ and $\theta = \pi$. Those extreme values are:

$$\text{Newton:} \quad \Omega \frac{(1 \pm e)^2}{(1-e^2)^{\frac{3}{2}}} \qquad \text{Ptolemy:} \quad \Omega \frac{1 \pm \frac{CE}{R}}{1 \mp \frac{TC}{R}}$$

Equating the two models and expanding $\frac{(1 \pm e)^2}{(1-e^2)^{\frac{3}{2}}} \approx 1 \pm 2e + \frac{5}{2}e^2$, we obtain the same result (8)[5].

# Conclusion

Optimizing (in the sense described) the shifts from Earth to the center of the deferent and from that center to the equant in Ptolemy's model produces a curious result: those two distances should be in ratio 5/3, at variance with Ptolemy's assumptions. This allows the best possible approximation of true elliptic motion to be precise to the second order in eccentricity, rather than to the first. This result is *universal*. It does not depend on the eccentricity of the true elliptical orbit: both shifts grow proportionally to that eccentricity, and remain in the same ratio. Of course, the eccentricity is still assumed to be much smaller than 1.

Strictly speaking, to achieve the second-order precision in the eccentricities of both Earth and the planet, we would ask Ptolemy to shift the center of the circle and the equant not just in the deferent, but also in the epicycle, and make the ratio of the shifts 5/3 in both. If we are not willing to introduce any new parameters into his model and rather want to tune the old ones, the success will be partial, but we can still achieve a substantial improvement in precision. Optimizing the deferent separately from the epicycle makes sense because, as we explained above, the epicycle motion is both smaller and much more uniform and circular than the deferent motion (with the notorious exception of Mercury). It actually makes sense to consider its irregularities as the next order of precision, that is, account for them in the first order when we account for irregularities in the deferent motion in the second order, and so on. Lastly, let us not forget the case of the Sun, which does not have an epicycle. In this case our optimization procedure is complete.

---

[5] Tycho Brahe found the ratio 5/3 empirically for the orbit of Mars.

What historical significance may these facts have? Below we present some thoughts about this, obviously, with a caveat that they are speculative in nature.

Most likely, Ptolemy did not optimize his model further because he did not see the need. He had a limited amount of (often low quality) observational data and wanted to restrict himself to the least sufficient modifications. But in the centuries after him, when more observational data were amassed, the deficiencies of Ptolemy's model became exposed. Deviations from the old astronomical tables also became very noticeable, due to error accumulation over time. All this eventually led Kepler to reject Ptolemy's model. But if he could achieve his ends by modifying the model, rather than by rejecting it, perhaps he would have done so, at least for some time. Kepler did not "prove" his laws. He showed only that they are compatible with observations. Indeed, proving the elliptical shape of the orbits based only on angular data is impossible. And Kepler would not have been driven towards his elliptical orbits if the data was explained better within Ptolemy's model, so that it would seem more natural to continue developing the model, perhaps by adding more epicycles. Therefore, it appears that the sub-optimal nature of Ptolemy's construction played an important role in the development of science: without it, the discovery and acceptance of Kepler's laws might have been postponed by a few decades. A term longer than that seems unlikely because of the creation of Newton's mechanics towards the end of seventeenth century. With it, the ability of a model to predict correct measurements was no longer the only criterion of validity: in Newton's mechanics, Ptolemy's planetary motion seems highly unnatural, even inexplicable, and this alone would discredit the model.

Trying to perturb his perfect-looking circular model to account for eccentricities, Ptolemy faced a choice: diminishing perfection was painful for an ancient Greek scientist, and he had to decide which way of doing it would minimize the pain. In the end, he chose to shift the center of his deferent circle and the equant. But he could, conceivably, choose to change that circle into an ellipse – Greeks knew conical sections and considered them beautiful, albeit less perfect than a circle. On the other hand, Kepler decided to depart completely from Ptolemy's model when, in principle, he could stay within it and just modify it. It is a curious thought that Ptolemy could have anticipated Kepler's model, and Kepler could have kept Ptolemy's model, and that both men's choices were, in the end of the day, esthetic.